\documentclass[10pt,aip,cha,twocolumn,showpacs,a4paper,floatfix]{revtex4-1}
\usepackage[dvips]{graphicx}
\usepackage[active]{srcltx}

\begin{document}
\bibliographystyle{plain}
\title{Chimeras in a Network of Three Oscillator Populations with Varying Network Topology}
\author{\firstname Erik A. \surname Martens}
\email{erik.martens@ds.mpg.de} 
\affiliation{
Max Planck Research Group for Biological Physics and Evolutionary Dynamics, 
Max Planck Institute for Dynamics and Selforganization, G\"{o}ttingen 37073, Germany}
\altaffiliation{Department of Theoretical \& Applied Mechanics, Cornell University, Ithaca 14853, USA}

\begin{abstract}
We study a network of three populations of coupled phase oscillators with identical frequencies. The populations interact nonlocally, in the sense that all oscillators are coupled to one another, but more weakly to those in neighboring populations than to those in their own population. Using this system as a model system, we discuss for the first time the influence of network topology on the existence of so called chimera states. In this context, the network with three populations represents an interesting case because the populations may either be connected as a triangle, or as a chain, thereby representing the simplest discrete network of either a ring or a line segment of oscillator populations. We introduce a special parameter that allows us to study the effect of breaking the triangular network structure, and to vary the network symmetry continuously such that it becomes more and more chain-like. By showing that chimera states only exist for a bounded set of parameter values we demonstrate that their existence depends strongly on the underlying network structures. We conclude that chimeras exist on networks with a chain-like character, which indicates that it might be possible to observe chimeras on a continuous line segment of oscillators.
\end{abstract}

\maketitle
\begin{quotation}
Collective synchronization of coupled oscillators is a problem of fundamental importance, and occurs in a wide range of systems including Josephson junctions arrays, circadian pacemaker cells in the brain or the metabolism of yeast cells~\cite{wiesenfeld1994, pikovsky2001, strogatz2003}. 
The Kuramoto model has been studied under the influence of global (all-to-all) and local (nearest neighbor) coupling in great detail.~\cite{kuramoto1984owt, acebron2005} The intermediate case of {\em nonlocal coupling} where the coupling strength decays with distance in a network was first investigated by Kuramoto \emph{et al.}.~\cite{kuramoto1995nto} In 2002 they observed a remarkable novel state where a population of \emph{identical} oscillators splits into two subpopulations, one being synchronized and the other desynchronized, which is called \emph{chimera state}.~\cite{kuramoto2002cca,battogtokh2002cin}
Since then, several studies have been concerned with its bifurcation behavior and its emergence under the aspects of heterogeneous oscillator frequencies or delayed coupling.~\cite{laing2009csh,sethia2005ccs,omelchenko2008cnl,abrams2008smc} Chimeras have been observed on a variety of network structures such as rings, networks with two and three oscillator populations, and 2D lattices in the shape of spiral waves.~\cite{abrams2004csc,abrams2004cro,martens2009cec,shima2004rsw,martens2010swc}
A natural question arises: which network structures allow for the existence of chimeras?~\cite{motter2010}
Here we determine for the first time limits for the existence of chimeras in a simple network of three oscillator populations~\cite{martens2009cec} as we vary the nature of the nonlocal coupling amongst the populations. 
\end{quotation}

\section{Introduction}\label{sec:introduction2}

While studying a continuum of identical oscillators on a ring with nonlocal coupling, Kuramoto \emph{et al.}~\cite{battogtokh2002cin} discovered a remarkable state where the population of oscillators splits into two subpopulations, where one is synchronized and the other is desynchronized, known as a \emph{chimera state}. Since then several groups have explored the nonlinear dynamics of chimeras.~\cite{battogtokh2002cin,abrams2004cro,abrams2004csc,shima2004rsw,kawamura2007ciw,kawamura2007hsn,pikovsky2008qpc,abrams2008smc,laing2009csh,omelchenko2008cnl,martens2010swc,bordyugov2010}
Their emergence on the ring was first analyzed by Abrams and Strogatz,~\cite{abrams2004cro,abrams2004csc} who found that chimera states are born through a saddle node bifurcation. 
The state exists in systems with various network structures; for instance, 
Shima and Kuramoto~\cite{shima2004rsw} showed that chimeras also exist on 2D lattices with free boundaries in the shape of spiral waves. Other authors have been studying its appearance on variations of the Kuramoto model such as systems with delayed coupling.~\cite{omelchenko2008cnl,sethia2005ccs}
 
While various aspects about the emergence of chimeras have been addressed, the question of how the underlying network structure affects their existence has not been addressed systematically. We make a first step in this direction by studying a network of three oscillator populations with nonlocal coupling. A network with three nodes may either be arranged as a triangle or as a chain, and therefore represents the simplest case of a discrete network with ring-like or chain-like character, respectively; the network structure will be determined by the nature of the coupling, as we explain later. The case of the triangular network has already been discussed in a previous study~\cite{martens2009cec} where the author shows that two stable chimera states may coexist. In this study, we shall introduce a new parameter that allows us to break the rotational invariance inherent to the triangle and to vary the network structure continuously such that it becomes more and more chain-like. Thereby we seek to find parameter limits for the existence of chimera states.

We consider the case of infinitely many oscillators. This case is often considered a valid approximation, and using a recently developed method by Ott and Antonsen,~\cite{ott2008ldb,ott2009lte} it enables us to reduce this infinitely dimensional problem to a set of a finite ordinary differential equations.~\cite{martens2009cec,abrams2008smc}

The article is structured as follows. In Section II we provide the definition of the system under investigation and explain how we intend to vary the character of the network structure by introduction of a special parameter. 
We then state the equations resulting from applying the Ott-Antonsen method  and consider special symmetries that allow for chimeras. 
The analysis of the chimera states and their bifurcation scenarios follows in Section~\ref{sec:analysis2}, where we analyze how the chimera states cease to exist as we vary the network structure using the above mentioned parameter. 
Section~\ref{sec:discussion2} summarizes our findings.

\section{Governing equations}\label{sec:governing equations2}
The governing equations are given by
\begin{eqnarray}\label{eq:goveqns2}
	\frac{d}{dt}\theta_i^{\sigma}&=&\omega + \sum_{\sigma'=1}^3 \frac{K_{\sigma\sigma'}}{N_{\sigma'}}\sum_{j=1}^{N_{\sigma'}} \sin{(\theta_j^{\sigma'} -\theta_i^{\sigma}  - \alpha)},
\end{eqnarray}
where the phases of the oscillators are defined by $\theta$; $i$ denotes the individual oscillators belonging to the population with index $\sigma\in\{1,2,3\}$, each of which has $N_{\sigma}$ oscillators; parameter $\alpha$ changes the character of the phase attraction.

\begin{figure*}[ht!]
\vspace{.5cm}
	\begin{center}
	\includegraphics[width=.95\textwidth]{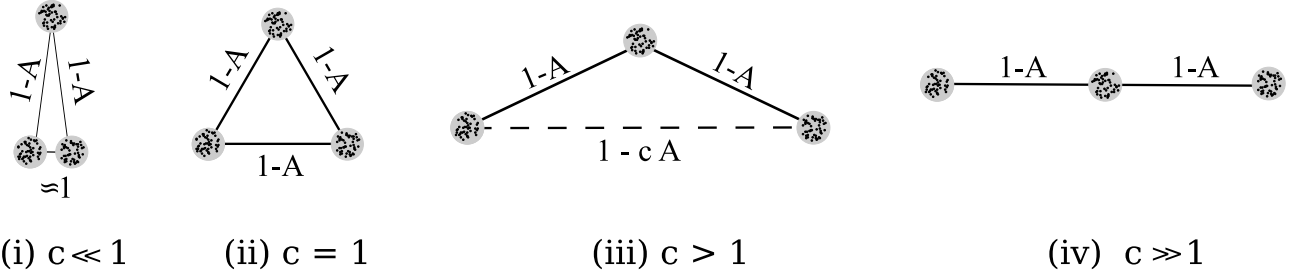}\
	\end{center}
\caption {Resulting network structures by varying parameter $c$. The gray disks symbolize populations, inhabited by individual oscillators symbolized by black dots. Their bidirectional coupling is represented by black lines.  Each population has a self-coupling of unit strength $1$. The population in what becomes the center for $c\neq1$ is coupled to the neighboring populations with strength $1-A$; the populations to the left and right are coupled with strength $1-cA$. The case of a triangular network is obtained for $c=1$; the character of the network has chain-like character for $c>1$.}
\label{fig:networks}
\end{figure*}

The coupling kernel $K_{\sigma\sigma'}$ describes the coupling strength between populations $\sigma$ and $\sigma'$ and is assumed to decay with increasing separation between the populations on the network. Within a population, the oscillators interact with strength $K_{\sigma\sigma'}=1$. Neighboring populations couple more weakly, with strength $1-A$ or $1-c\,A$, as displayed in Fig.~\ref{fig:networks}. We then have
\begin{equation}\label{eq:kernel2}
	K_{\sigma\sigma'} = \left(\begin{array}{lcr}
			1    & 1-A & 1-c\,A\\
                        1-A  & 1   & 1-A\\
			1-c\,A & 1-A & 1\
                       \end{array}
\right). 
\end{equation}
In the case where $A=0$ we retrieve the case of a globally coupled network. Thus $A$ may be thought of how 'far' we are away from global coupling. The introduction of the parameter $c$ in (\ref{eq:kernel2}) generalizes our previous work~\cite{martens2009cec}: for $c=1$, one obtains the case of a triangular network with the same rotational symmetry as a continuum of oscillators on a ring, studied by.~\cite{battogtokh2002cin,abrams2004cro,abrams2004csc,abrams2008smc}
If we let $c\neq1$ this rotational symmetry is broken; in particular, for $c>1$, the 'left' and 'right' populations interact less strongly with one another than they interact with what turns out to be the center population (see Fig.~\ref{fig:networks} (iii)). Hence parameter $c$ controls what one may conceive as the distance between the outer left and right population. Thus for $c>1$, the network attains a chain-like character. To facilitate comparison with earlier work, we focus on positive coupling only and impose the constraints 
\begin{eqnarray}
 	c\cdot A\leq 1,\\
	A \leq 1.\
\end{eqnarray}
Our special interest lies in the case where $c>1$ because of its close relation to chain-like structures. However, we relax this constraint during the subsequent analysis in favor of a deeper understanding of the bifurcation structure.

\subsection{Reduced equations and symmetry manifolds}
We summarize how we obtain a reduced set of governing equations based on the Ott-Antonsen method~\cite{ott2008ldb}.
In order to analyze the behavior of (\ref{eq:goveqns2}), we consider the limit of infinitely large populations, i.e. $N_{\sigma}\rightarrow \infty$. In this limit, one may describe the dynamics in terms of the oscillator density distribution $f^{\sigma}(\theta,t)$, which evolves according to the continuity equation
\begin{eqnarray}
 \frac{\partial f^{\sigma}}{\partial t} + \frac{\partial}{\partial \theta}(f^{\sigma}v^{\sigma}) &=& 0.\
\end{eqnarray}
The velocity of the oscillators in this limit is given by
\begin{eqnarray}\nonumber
 	v^{\sigma}&=&\omega + \sum_{\sigma'=1}^3 K_{\sigma\sigma'}\int_0^{2\pi} \sin{(\theta'-\theta-\alpha)}f^{\sigma'}(\theta',t)d\theta', \\
\end{eqnarray}
and a complex order parameter is introduced,
\begin{eqnarray}\label{eq:cop2}
 z^{\sigma}(t) &=& \sum_{\sigma'=1}^3 K_{\sigma\sigma'}\int_0^{2\pi}e^{i\theta'}f(\theta',t)\,d\theta',\
\end{eqnarray}
defining an average over all oscillator phases; it is therefore referred to as \emph{weighted or global order parameter}. 
In this formulation of the problem, we are able reduce the problem to a finite set of equations using a method recently developed by Ott and Antonsen.~\cite{ott2008ldb} 
Following Ott and Antonsen, we restrict our attention to a special class of density functions in the form of 
\begin{eqnarray}\label{eq:poissonkernel}
 	f^{\sigma}(\theta,t)&=& \frac{1}{2\pi}\left\{1+\left[\sum_{k= 1}^{\infty}(a_{\sigma}(t)e^{i\theta})^k + c.c.\right]\right\},\
\end{eqnarray}
where c.c. is the complex conjugate of the expression under the sum. The application of this ansatz ultimately leads to a set of ordinary differential equations.
How  this reduction method is performed on the problem of oscillator populations and leads to a finite set of equations has been reported in detail elsewhere.~\cite{abrams2008smc,ott2009lte,martens2009cec} We skip the full derivation and state the resulting reduced equations:
\begin{eqnarray}\label{eq:lodimbeta2}
 \dot \rho_{\sigma}&=& \frac{1-\rho_{\sigma}^2}{2}\sum_{\sigma'=1}^{3}K_{\sigma\sigma'}\rho_{\sigma'}\sin{(\phi_{\sigma'}-\phi_{\sigma}+\beta)}, \\
\dot \phi_{\sigma}&=&\omega- \frac{1+\rho_{\sigma}^2}{2\rho_{\sigma}}\sum_{\sigma'=1}^{3}K_{\sigma\sigma'}\rho_{\sigma'}\cos{(\phi_{\sigma'}-\phi_{\sigma}+\beta)},\
\end{eqnarray}
where 
\begin{eqnarray}
	\beta&=&\pi/2-\alpha,\\
	a_{\sigma}&=&\rho_{\sigma} \,e^{-i \phi_{\sigma}},\\
	\sigma &\in& \{1,2,3\}.\
\end{eqnarray}
The variable $a^*_{\sigma}$ takes account for the \emph{local} order for each population and found to be related to the global order parameter via
\begin{eqnarray}
 z_{\sigma}&=& \sum^3_{\sigma'=1}K_{\sigma\sigma'}a^*_{\sigma'}(t).\
\end{eqnarray}
The dynamics of the oscillators is described by two variables per population: $\rho_{\sigma}$  describes their degree of synchronization and $\phi_{\sigma}$ represents the average phase of the oscillators in each population. 

Before analyzing these equations, we restrict our attention to specific symmetry manifolds.  Perfectly synchronized populations have $\rho_{\sigma}=1$ and desynchronized populations have $\rho_{\sigma}<1$; the latter consists of oscillators that drift relative to one another and to the synchronized populations.
Let $S$ and $D$ denote synchronized and desynchronized populations, respectively. 
In the triangular network, due to its rotational invariance, we could only  distinguish two chimera states, namely $SDS$ (sync-drift-sync) and $DSD$ (drift-sync-drift). The situation is different here, and other chimeras are possible in the case of $c\neq1$: $SDD$, $SSD$ and their symmetry-equivalent reflections across the center population. These states are however not the focus of this study and are excluded in our analysis.

The $SDS$ state is defined as $\rho_1 = \rho_3 = 1$ and $\rho\equiv\rho_2<1$, whereas the  $DSD$ state has $\rho\equiv\rho_1 = \rho_3<1$ and $\rho_2=1$. Because our coupling kernel (\ref{eq:kernel2}) is symmetric, $\rho_1 = \rho_3$ also implies $\phi_1=\phi_3$; hence populations $1$ and $3$ are phase-locked (this holds true also if $c\neq1$.). The phase difference of the angular order parameter between the synchronized and desynchronized states  is defined by
\begin{eqnarray}
	\psi=\phi_1-\phi_2=\phi_3-\phi_2.\
\end{eqnarray}
Applying these symmetry assumptions to (\ref{eq:lodimbeta2}) and  substituting the coupling kernel defined in (\ref{eq:kernel2}), we obtain the equations describing the $SDS$ states
\begin{eqnarray}\label{eq:SDShybrid2}	\nonumber	
	\dot \rho&=&  \frac{1-\rho^2}{2}\left[2(1-A)\sin{(\psi+\beta)}+\rho\sin{\beta}\right],\\\nonumber
	\dot \psi&=& -(2-cA)\cos{\beta} - (1-A)\rho\cos{(-\psi+\beta)}\\
	&+&\frac{1+\rho^2}{2\rho}\left[2(1-A)\cos{(\psi+\beta)} +\rho\cos{\beta}\right],\
\end{eqnarray}
and the $DSD$ states
\begin{eqnarray}\label{eq:DSDhybrid2}\nonumber
	\dot \rho&=&  \frac{1-\rho^2}{2}\left[(2-cA)\rho\sin{\beta} + (1-A)\sin{(-\psi+\beta)}\right],\\\nonumber
	\dot \psi&=& -\frac{1+\rho^2}{2\rho}\left[(2-cA)\rho\cos{\beta} + (1-A)\cos{(-\psi+\beta)}\right]\\
	&+&2(1-A)\rho\cos{(\psi+\beta)} + \cos{\beta}.\
\end{eqnarray}
Fixed points of (\ref{eq:SDShybrid2}) and (\ref{eq:DSDhybrid2}) correspond to phase-locked solutions of the original system.
The reduced equations (\ref{eq:lodimbeta2}) (after transformation into the co-rotating frame, i.e. $\phi_{\sigma}\rightarrow \phi_{\sigma}+\omega t$), and (\ref{eq:SDShybrid2},\ref{eq:DSDhybrid2}) share the property of being invariant under the following time-reversibility transformations:
\begin{eqnarray}\label{eq:symmetry1}
	(\beta,t,\psi)&\rightarrow&(-\beta,-t,-\psi),\\\label{eq:symmetry2}
	(\beta,t)&\rightarrow&(\beta+\pi,-t).\
\end{eqnarray}
The bifurcation structures discussed in the next section repeat themselves accordingly in the $(\beta,A)$-plane.

In conclusion, we have reduced the governing equations (\ref{eq:goveqns2}) to a low dimensional system for the local order parameters, and we have cast our problem into a two dimensional system represented by Eqs.~(\ref{eq:SDShybrid2}) and (\ref{eq:DSDhybrid2}). This enables us now to study the problem in the phase plane.

\section{Bifurcation behavior near the triangular structure}\label{sec:analysis2}
\begin{figure*}[!ht]
	\begin{center}
	 \includegraphics[width=.95\textwidth]{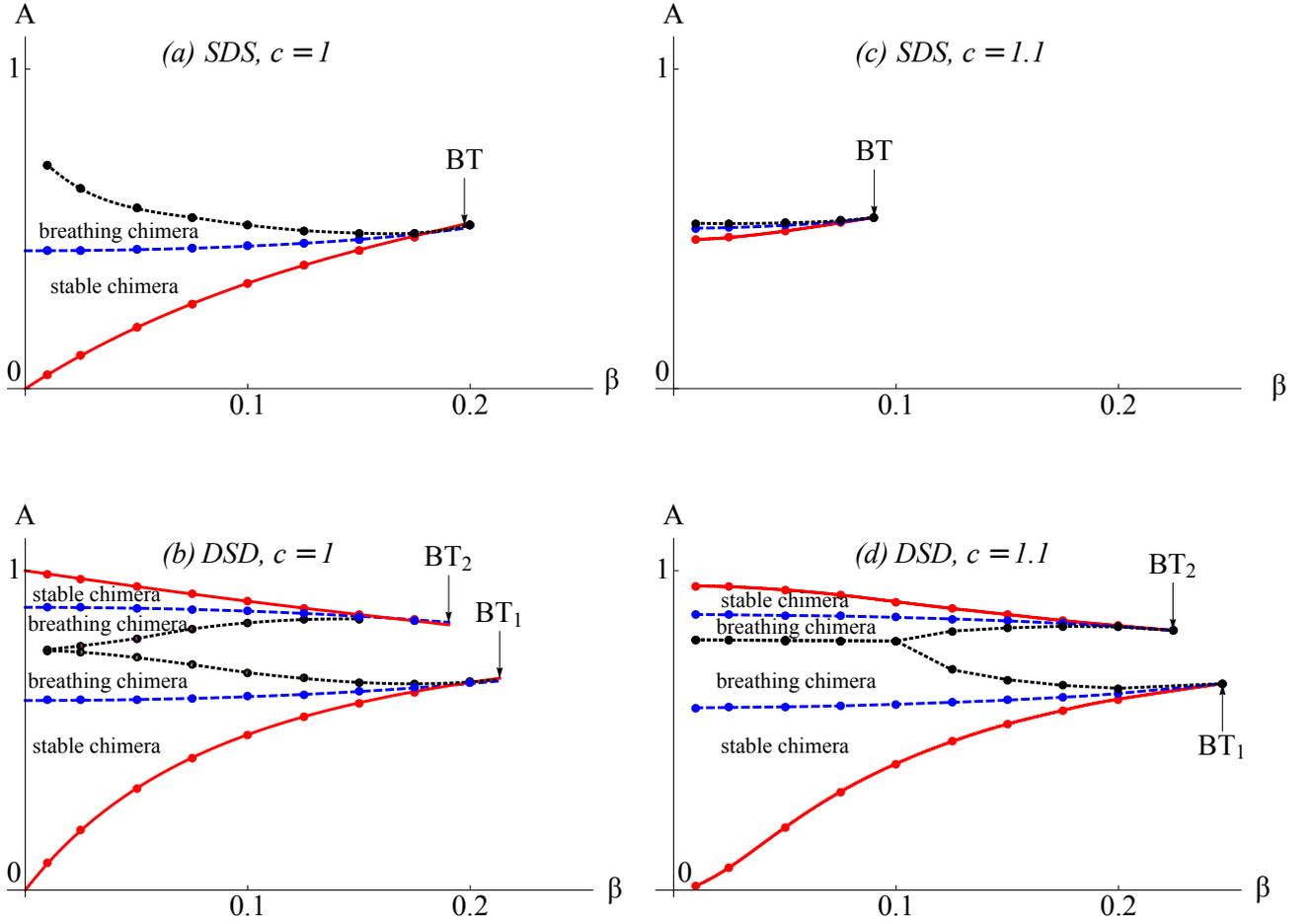}
	\end{center}
\caption{Effect of breaking the rotational symmetry on the bifurcation diagram for the $SDS$ and $DSD$ symmetries.  The triangular case~\cite{martens2009cec} with $c=1$ is shown in the left column in (a) and (c) for comparison with the case of broken symmetry ($c=1.1$) in the right column in (b) and (d). The displayed curves are: the saddle-node curve (solid red), the Hopf curve (dashed blue), and the homoclinic curve (dotted black). Dots mark the bifurcation points obtained by inspection of the phase plane. The homoclinic curve is an interpolation based on these points,  whereas the solid curves were obtained analytically.}
\label{fig:BD_hybridc1}
\end{figure*}

We briefly review the findings~\cite{martens2009cec} for the triangular coupling ($c=1$, before we study the symmetry breaking case ($c\neq 1$) in more detail. The associated bifurcation diagrams are shown in Fig.~\ref{fig:BD_hybridc1}, where we compare the triangular case (left) with the symmetry breaking case (right). We first consider the case of $SDS$ symmetry. 
We keep the parameter $\beta$ constant while increasing the value of $A$ step by step:
close to global coupling, i.e. for small values of $A$, we only observe the always present in-phase $SSS$ solution.  
As we increase $A$, a stable chimera is born in  a saddle-node bifurcation. 
One step further, the chimera loses its stability through a Hopf bifurcation and a limit cycle is born, which corresponds to a so-called \emph{breathing chimera} with an oscillating order parameter. 
Increasing $A$ further, we observe how the limit cycle grows and eventually collides with the saddle and is destroyed in a homoclinic bifurcation. The saddle-node curve, Hopf and homoclinic bifurcation curves all intersect in the Bogdanov-Takens (BT) point of codimension two.
The $DSD$ symmetry exhibits a similar bifurcation structure: for small $A$, the scenario is identical to the one seen for the $SDS$ symmetry; however for larger $A$, the scenario is repeated in \emph{reversed order} (with increasing values of $A$), as shown in Fig.~\ref{fig:BD_hybridc1}.

We consider now the symmetry breaking case relatively close to the triangular symmetry with $c=1.1$. The bifurcation diagrams are obtained by inspection of phase portraits (large dots in the figure). 
Saddle node and Hopf curves may be calculated analytically by simultaneously solving the fixed point equations (\ref{eq:SDShybrid2}) and (\ref{eq:DSDhybrid2}) with either the saddle node condition,
\begin{eqnarray}\label{eq:SN}
 \det{(J)}&=&0\
\end{eqnarray}
or the Hopf condition,
\begin{eqnarray}\label{eq:Hopf}
 \textrm{tr}{(J)}&=&0\ \quad\textrm{and} \quad \det{(J)}>0,\
\end{eqnarray}
where $J$ is the Jacobian of (\ref{eq:SDShybrid2}) and (\ref{eq:DSDhybrid2}). The resulting equations are solved using a series approach with $\beta$ as a bifurcation parameter, similar to the method described in detail in earlier work,~\cite{martens2009cec} however by assigning a desired value to  $c$ before computing the series coefficients. The perturbation series for $A=A(\beta)$ yield no further insight and we omit them here for brevity.
We observe a similar bifurcation scenario as before in the triangular case, but there is an important qualitative change: the parameter region allowing for $SDS$  chimeras (above) has shrunk in parameter space, whereas the attractor region of $DSD$ chimeras (below) has grown. 
Interestingly, it looks as if the homoclinic curves for the $DSD$ chimera now coincide for a wide range of $\beta$ values. Unfortunately we could not confirm this behavior analytically, as we were unable to determine the Melnikov integrals leading to homoclinic conditions, but the phenomenon is not essential for the matter of this study. 

Moreover, the Bogdanov-Takens (BT) points have moved to the left and right, respectively, according to the shrinking or growing of the regions of existence for chimeras. 
We can check this by determining the locus of the BT points for $c=1.1$. We numerically solve the fixed point equations implied by Eqs.~(\ref{eq:SDShybrid2}) and (\ref{eq:DSDhybrid2}) simultaneously with the saddle node (\ref{eq:SN}) and Hopf condition (\ref{eq:Hopf}). For $c=1.1$ we find the following critical points for the $SDS$ symmetry,
\begin{eqnarray}
	(\beta,A)_{SDS}   &\approx& (0.0902, 0.5361),\
\end{eqnarray}
and for the $DSD$ symmetry,
\begin{eqnarray}
	(\beta,A)_{DSD,1} &\approx& (0.2467, 0.6466)\\
	(\beta,A)_{DSD,2} &\approx& (0.2244, 0.8132),\ 
\end{eqnarray}
which are denoted by arrows in Fig.~\ref{fig:BD_hybridc1}. 
Comparison of these numbers with our previous results~\cite{martens2009cec} for $c=1$ confirms our observation that the BT points have shifted to the left and right, respectively. We proceed with finding analytical expressions describing how these points move in parameter space and put limits on the existence of chimeras.

\section{Limits of existence for chimeras}
\begin{figure}[ht!]
	\begin{center}
		\includegraphics[width=.45\textwidth]{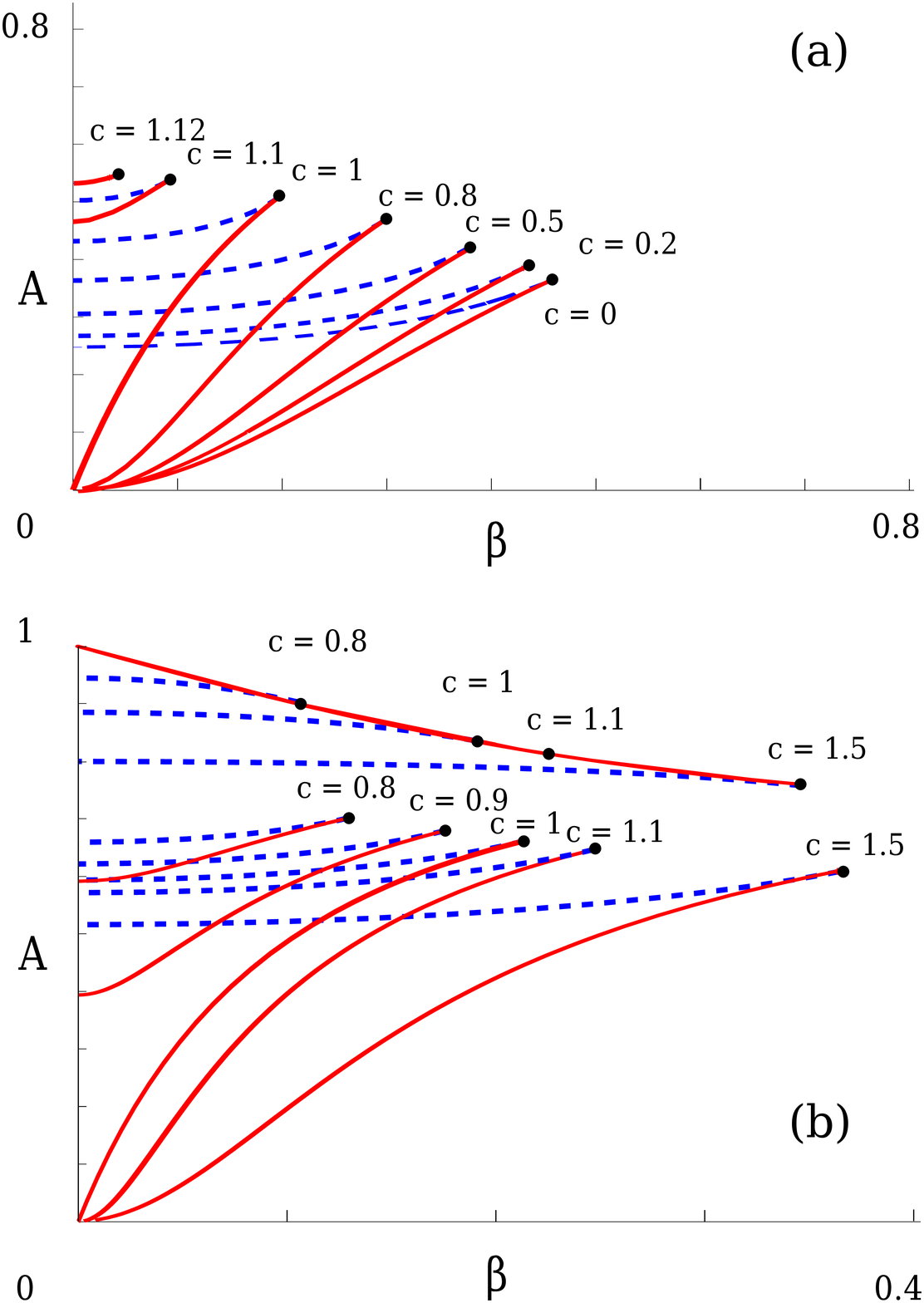}
	\end{center}
\caption{Bifurcation diagram for the $SDS$ chimera (above) and the two $DSD$ chimeras (below) in the $(\beta,A)$-plane for a range of $c$ values. Saddle node (solid red) and Hopf curves (dashed blue) are shown. The curves related to the $SDS$ chimera collapse onto the $\beta$-axis at $c\approx1.123$; conversely, the two $DSD$ curves collapse on the axis at $c\approx0.6778$ and $c=2/3$, respectively. It is seen that the Bogdanov-Takens point (black dots) of the upper $DSD$ chimera follows the associated saddle node curve as we vary $c$.}
\label{fig:BD_SDS_cmany}
\end{figure}

We study now more generally what happens to the chimeras as we vary the values of $c$. Bifurcation diagrams for $SDS$ and $DSD$ chimeras were obtained by numerical continuation for a range of $c$ values (Fig.~\ref{fig:BD_SDS_cmany}). 
The behavior described above that the attracting region for the $SDS$ chimeras shrinks and the region for $DSD$ chimeras grows with increasing values of $c$, is now confirmed for a larger range of parameter values; conversely, as we consider the case $c<1$, the opposite holds true.
Regions of existence for chimeras disappear completely at the point where the BT point touches the $\beta$-axis: this is where the saddle node, Hopf and homoclinic curves collapse in a single point.

We calculate the critical $c$ values where this occurs.
The $\beta$-axis represents a singular limit and simply substituting $\beta=0$ into the equations defining the BT points (saddle node (\ref{eq:SN}) and Hopf condition (\ref{eq:Hopf})), leads nowhere; however we can calculate $c_{crit}$ by using a perturbative approach. We use $\beta$ as a perturbation parameter in  the limit $\beta\rightarrow 0$. Inspecting the phase portrait of (\ref{eq:SDShybrid2}) we find that the $SDS$ chimera is located near $\psi=0$; the correct perturbation ansatz is
\begin{eqnarray}
	\psi =  \psi_1\,\beta+O(\beta^2).\
\end{eqnarray}
Solving the resulting equations at first order of $\beta$, we find that the critical point for the $SDS$ symmetry is located at
\begin{eqnarray}\label{eq:ccrit1}\nonumber
	c_{crit} &=& \frac{1}{12} \left(15-\sqrt{33}+2 \sqrt{ 6\sqrt{33}-30}\right) \approx 1.12356,\\\nonumber
	A_{crit} &=&1-\frac{\sqrt{2}}{32} \left(\sqrt{33} -1\right)^{3/2} \approx 0.54327,\\
	\rho_{crit} &=&\sqrt{\frac{\sqrt{33}-1}{8}}\approx 0.770111.\
\end{eqnarray}
By the same token, for the $DSD$ chimera associated with \emph{small} values of $A$ (lower $DSD$ chimera), we have
\begin{eqnarray}\label{eq:ccrit2}\nonumber
	c_{crit} &=& \left(4774-350 \sqrt{57}
	    +\left(473 \sqrt{7} -53 \sqrt{399}\right)\right.\\
	    &&\times\left.\sqrt{(1+\sqrt{57}}\right) / 3976\approx 0.677869,\\\nonumber
	A_{crit} &=& 1+\sqrt{1+\sqrt{57}}\\
	&&\times\left(11\sqrt{399} -151 \sqrt{7}\right)/1988 \approx 0.735569,\\
	\rho_{crit}&=& \frac{1}{2} \sqrt{\frac{1}{7} \left(1+\sqrt{57}\right)} \approx 0.552586.\ 
\end{eqnarray}

Observing Fig.~\ref{fig:BD_SDS_cmany}, we notice that the saddle node curves of the upper $DSD$ states coincide for all values of $c$; hence the associated set of BT points  must also be located on this curve. Following the motion of the BT point as $c$ decreases, we suspect that $c$ becomes critical exactly when the BT point has reached $A=1$. (This looks like a limit of a higher order singularity than previously; indeed it is much harder to see in phase portraits at which angle $\psi$ the BT point detaches from the $\beta$-axis.) This time the correct perturbation ansatz turns out to be
\begin{eqnarray}
	\psi = \frac{\pi}{2} + \psi_1\,\beta+O(\beta^2).\
\end{eqnarray}
Solving again to first order in $\beta$ results in
\begin{eqnarray}\label{eq:ccrit3}\nonumber
	c_{crit}	&=& \frac{2}{3},  \\\nonumber
 	A_{crit}	&=& 1,  \\
	\rho_{crit}	&=& \sqrt{\frac{1}{2}}\,.\
\end{eqnarray}
We conclude that $SDS$ chimeras exist for parameter values $c\leq1.12356$, and $DSD$ chimera for $c\geq0.677869$ and for $c\geq2/3$, respectively (we shall discuss below that these limits for the existence of stable chimeras hold if $\beta>0$).

\begin{figure}[!ht]
	\begin{center}
		\includegraphics[width=0.45\textwidth]{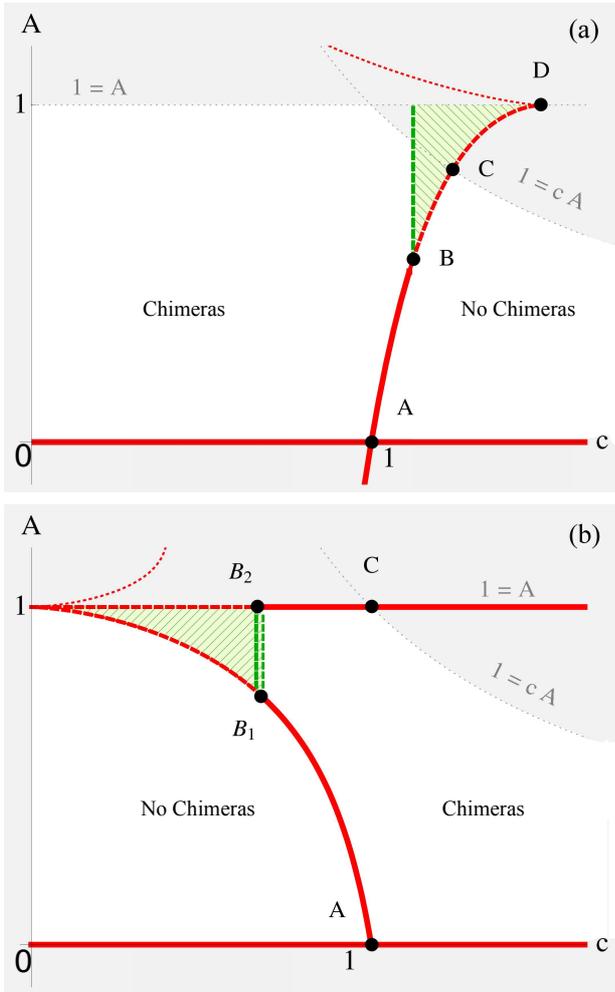}\
	\end{center}
\caption{
	Boundaries for the occurrence of saddle node transitions in the $(c,A)$-plane (i.e. saddle node curves at $\beta=0$), shown for the $SDS$ symmetry (a) and the two $DSD$ symmetries in (b) ($A=1$ is the boundary for the second $DSD$ state seen in the upper part in Fig.~\ref{fig:BD_SDS_cmany} b)). The regions shaded in gray either have negative coupling or have coupling without chain-like character, as  explained in the text. Throughout the regions labeled with {\em No Chimera}, we find no chimeras. The black dots indicate points of special interest: 
	{\bf A}: $A=(1,0)$ is where the saddle node curves detach from the origin $(\beta,A)$-plane. 
	{\bf B}: are the points for which  the BT points collide with the $\beta$-axis leading 
	to the annihilation of the chimera state; $B\approx(1.123,0.543)$, $B_1\approx(0.677,0.735)$ and $B_2=(2/3,1)$. 
	{\bf C}: intersection with boundaries of positive coupling; $C_{SDS}=((27+\sqrt{27})/26,26(27+\sqrt{27}))$ and $C_{DSD}=(1,1)$. 
	{\bf D}: intersection of the saddle-node boundary with $1=A$; $D=(3/2,1)$.
	The regions where chimera exist for $\beta<0$, which are stable within the symmetry manifolds $SDS$ and $DSD$, are hatched green. 
}
\label{fig:Acdiagram}
\end{figure}

Another interesting phenomenon is seen in Fig.~\ref{fig:BD_SDS_cmany}: the saddle node curves of the $SDS$ chimera pass through the origin $(\beta,A)=(0,0)$ only if $c<1$, but detach from the origin as soon as $c>1$. 
The analogous behavior is seen for the $DSD$ chimera, however for the converse case where  $c<1$.  
Thus having passed $c=1$ and approaching $c_{crit}$ implies the shrinking of the chimera attractor regions not only in the direction of the $\beta$-axis, but also along the direction of the $A$-axis.
To shed more light on this behavior, we calculate the locus of the saddle node transition $A=A(c)$ for $\beta\rightarrow 0$. 
These curves are this time simply obtained by letting $\beta=0$ while solving the saddle node condition together with the fixed point conditions. 
For $SDS$, we have:
\begin{eqnarray}\label{eq:Ac_SDS1}
	A &=& \frac{3}{16 c^3} \left(12 c^2-9\pm\sqrt{3} \sqrt{(3-2 c)^3 (1+2 c)}\right), \\\label{eq:Ac_SDS2}
	A &=&0,\
\end{eqnarray}
and for $DSD$, we obtain the equations
\begin{eqnarray}\label{eq:Ac_DSD1}\nonumber
	27 (A-2) (A-1)^2 A 
	&=& A c \left[-54+18 A (6+c)\right.\\\nonumber
	&& + A^3 c \left(18+c^2\right)\\
	&& - \left. 2 A^2 (27+c (18+c))\right],\\
	  \label{eq:Ac_DSD2}
	A&=&1.\
\end{eqnarray}
The resulting curves are shown in red in Figs.~\ref{fig:Acdiagram} (a) and (b). 
Recall that we are only interested in regions representing positive coupling, and regions where the coupling decays with increasing distance on the network graph. 
The second and forth quadrants in the $(c,A)$-plane represent regions where the coupling matrix has a non-decaying character because of its entry with $(1-cA)>1$ in (\ref{eq:kernel2}), and has therefore no chain-like character. 
The entire lower half-plane has the same problem, but now due to the coupling kernel entry $(1-A)>1$.
In the quadrant that is left, we have to respect the constraints $A\leq 1$ and $1 \leq c\cdot A $; they are indicated by gray dotted curves. 
All these regions that are not of interest to us are shaded in gray.
The curves defined by (\ref{eq:Ac_SDS1})-(\ref{eq:Ac_DSD2}) reach into these forbidden areas and are related to interesting dynamics; its nature is beyond the scope of this study and would require a more in-depth analysis.

Special points of interest are labeled as follows: A is where the saddle node curve detaches from the $\beta$-axis, as shown in Fig.~\ref{fig:BD_SDS_cmany}; the various $\textrm{B}_i$'s represent the critical points where the BT points, together with their stable and breathing chimera regimes, collide with the $\beta$-axis, for $SDS$ and $DSD$ symmetries; C denotes the point where the 'upper' $DSD$ chimera boundary (\ref{eq:Ac_DSD2}) intersects with the $1=c\cdot A$ boundary; and finally, D denotes the analogous intersection with the $A=1$ boundary.

Note that the curves described by Eqs.~(\ref{eq:Ac_SDS1})-(\ref{eq:Ac_DSD2}) do not delineate complete regions of stability: they represent the boundary of lowest possible $A$ values where a saddle-node bifurcation may occur for at least one $\beta$ (this point always lies on the $\beta=0$ axis due to the observed monotonicity of the saddle-node curves). 
In other terms, chimeras that have been created through a saddle-node transition while increasing $A$ may lose their stability through a Hopf and get destroyed in a homoclinic bifurcation for larger values of $A$, depending on the value of $\beta$. Thus the curves should be understood as lower boundaries (or upper for $A=1$,  respectively, for the second $DSD$ state) where chimeras still exist. Accordingly in Fig.~\ref{fig:Acdiagram}, regions where chimeras cannot exist at all are labeled with \emph{No Chimera}, and regions where stable chimeras exist (for an appropriate choice of $\beta$) with \emph{Chimeras}.

Finally, we have a brief look at states with $\beta<0$ which so far have been neglected in previous studies.
The following applies for both $SDS$ and $DSD$ chimeras, but for simplicity let us consider only the $SDS$ symmetry.  
At point B, the BT point collides with the $\beta$-axis, and we expect not to see any chimeras beyond this point.
But for $c>c_{crit}$, after crossing the curve defined by $\overline {BD}$, one observes the creation of a saddle and a source with $\rho<1$, corresponding to an \emph{unstable} chimera. 
However, virtue to the symmetry in (\ref{eq:symmetry1}), $(\beta,\psi,t)\rightarrow(-\beta,-\psi,-t)$, a related stable node exists for negative $\beta$. 
Hence if we extend our study to include $\beta<0$, stable $SDS$ chimeras are also found for $c>c_{crit}$, as well as the two $DSD$ chimera states beyond their corresponding $c_{crit}$-values. The green hatched areas denote their regions of existence for $\beta<0$.
Still, they leave our domain of interest in Point C and completely cease to exist in D.
We note that no Hopf or homoclinic bifurcations occur in these regions: they appear to be completely annihilated in point $B$ (where the BT points collide with the $\beta$-axis).
Moreover, according to the symmetry defined by (\ref{eq:symmetry1}), there should also be instances of stable chimeras for $c<c_{crit}$ and $\beta<0$: the chimeras that are stable or breathers for $\beta>0$ are present with inverted stability, i.e. the source inside the limit cycle becomes a stable chimera for $\beta<0$ .

It is however conceivable that these states with $\beta<0$ are unstable towards perturbations in the transverse directions of the symmetry manifolds $SDS$ and $DSD$ (just as it seems to be the case for the $DSD$ in the upper part of the $(A,\beta)$-plane;~\cite{martens2009cec}) their occurrence is however beyond the scope of our interest and we leave this question for others to solve.

\section{Discussion}\label{sec:discussion2}
We have investigated the problem of three nonlocally coupled oscillator populations, thereby generalizing work on networks with triangular symmetry.~\cite{martens2009cec}
By introducing the structural parameter $c$, we were able to study the behavior of oscillator populations as we change the quality of the nonlocal coupling kernel (\ref{eq:kernel2}). These changes differ from the quality changes induced by parameter $A$, controlling how close we are to local or global coupling. Conversely, parameter $c$ controls the distance between the 'left' and 'right' populations, and we may distinguish the different qualities it tunes into as visualized in Fig.~\ref{fig:networks}: 
(i) for very small $c$, the left and right populations are very close, and they see the center as a 'satellite' population. In this limit, the presence of the $SDS$ chimera is dominating. One might argue that it corresponds -- in our terminology -- to the $SD$ chimera observed in two oscillator populations by Abrams {\em et al.}~\cite{abrams2008smc};
(ii) for $c=1$, we retrieve the triangular case with rotational invariance; 
(iii) for $c>1$ the network acts like a chain (discussed below) and 
(iv) for very large $c$, the outer populations will almost only sense each other's motions indirectly via their coupling to the center population, as long as $1\geq c\cdot A$, i.e. the coupling stays positive.

We determined lower and upper limits (\ref{eq:ccrit1}) for $c$, (\ref{eq:ccrit2}) and (\ref{eq:ccrit3}), where chimeras cease to exist.
We found that these limits are valid for $\beta>0$, but that stable chimera may exist (within other limits) beyond the critical $c$ values for $\beta<0$. 
Furthermore, we were able to find limiting curves $A=A(c)$ defining coupling parameter regions $(c,A)$ for which chimeras cannot exist at all.

We found that parameter $c$ needs to stay relatively close above $1$ to observe a {\em chain-like} $SDS$ chimera state. 
Regardless this closeness to the triangular symmetry $c=1$, any case with $c\neq1$ may be considered a valid instance of the discretization of a chain. For example, consider the case of a line-segment subject to a coupling kernel with exponential decay, $e^{-\kappa |x|}$; then we may choose the characteristic length scale $\kappa^{-1}$ of the kernel to match a desired value of $c$. It is therefore likely that chimeras may also exist for a continuum of oscillators on a line segment.

It is worth noting how the effects of the coupling parameters $A$ and $c$ are related to coupling kernels studied in other chimera systems. Abrams \emph{et al.}~\cite{abrams2004cro} discussed the emergence of chimeras on a continuum of oscillators on  a ring and thereby used a kernel $G(x) = 1+A \cos{(x)}$. Clearly, the effect of $A$ is the same as in our study by modulating the amplitude of the coupling. In an earlier study (also on a ring), Kuramoto and Battogtokh~\cite{kuramoto2002cca} used an exponential kernel $G(x) \sim \exp(-\kappa |x|)$ where $\kappa$ tunes the width of the nonlocal coupling. If space is discretized, the effect of varying this width is similar to controlling the 'distance' between the resulting oscillators populations, just like our parameter $c$ does it here. In this sense, our choice for the coupling unites both qualities.
In all systems using kernels with a modulation parameter $A$, it is found that chimeras emerge in the limit where $\beta=0$, where the system is nearly conservative. 
On the other hand, in a recent study~\cite{martens2010swc} we found that spiral wave chimeras in the plane - coupled via a Gaussian $G(x)=\exp{(- |x|^2)}$ - emerge from the different limit $\beta = \pi/2$ (i.e. near the gradient system limit). The precise origin for this adverse behavior or how the emergence of chimera states in these quite distinct limits is related to the different types of coupling remains to be solved. 

The issue of which classes of networks allow for chimera states to exist \emph{in general} is an open problem. A related matter is the characterization of the basins of attraction leading to chimera and globally synchronized states, a question raised just recently.~\cite{motter2010} For future research, we suggest to do more studies along these lines.

\section{Acknowledgments}
This research was supported in part by NSF Grant No. DMS-0412757 and CCF 0835706. I would like to thank Steve Strogatz for helpful discussions and advice throughout the scope of this project, and F. Schittler-Neves, G. Bordyugov and A. Pikovsky for valuable discussions.

\end{document}